\documentstyle[11pt,aaspp4,graphicx]{article}

\begin{document}

\title{Chromatic Signatures in the Microlensing of GRB Afterglows}

\author{ 
Jonathan Granot\altaffilmark{1} 
and Abraham Loeb\altaffilmark{2}}

\altaffiltext{1}{Racah Institute, Hebrew University, Jerusalem 91904,
  Israel; jgranot@merger.fiz.huji.ac.il}

\altaffiltext{2}{Harvard-Smithsonian Center for Astrophysics, 60
  Garden Street, Cambridge, MA 02138; aloeb@cfa.harvard.edu}

\begin{abstract} 

We calculate the radial surface brightness profile of the image of a
Gamma-Ray-Burst (GRB) afterglow. The afterglow spectrum consists of several
power-law segments separated by breaks. The image profile changes
considerably across each of the spectral breaks. It also depends on the
density profile of the ambient medium into which the GRB fireball
propagates.  Gravitational microlensing by an intervening star can resolve
the afterglow image. We calculate the predicted magnification history of
GRB afterglows as a function of observed frequency and ambient medium
properties. We find that intensive monitoring of a microlensed afterglow
lightcurve can be used to reconstruct the parameters of the fireball and
its environment, and provide constraints on particle acceleration and
magnetic field amplification in relativistic blast waves.

\end{abstract}


\section{Introduction}

The fireball of a Gamma-Ray Burst (GRB) afterglow is predicted to appear on
the sky as a ring (in the optical band) or a disk (at low radio
frequencies) that expands laterally at a superluminal speed, $\sim \Gamma
c$, where $\Gamma\gg1$ is the Lorentz factor of the relativistic blast wave
which emits the afterglow radiation (Waxman 1997; Sari 1998; Panaitescu \&
M\'esz\'aros 1998; Granot, Piran, \& Sari 1999a,b). Days after the GRB
trigger, the physical radius of the afterglow image ($\sim$ the fireball
radius over $\Gamma$) translates to an angular size of order a
micro-arcsecond ($\mu$as) at a cosmological distance.  Coincidentally, this
image size is comparable to the Einstein angle of a solar mass lens at a
cosmological distance,
\begin{equation}
\theta_{\rm E}=\left({4GM_{\rm lens}\over c^2 D}\right)^{1/2}= 1.6
\left({M_{\rm lens}\over 1 M_\odot}\right)^{1/2}\left({D\over 10^{28}~{\rm
cm}}\right)^{-1/2}~\mu{\rm as},
\label{eq:1}
\end{equation}
where $M_{\rm lens}$ is the lens mass, $D\equiv {D_{\rm os}D_{\rm ol}/
D_{\rm ls}}$ is the ratio of the angular-diameter distances between the
observer and the source, the observer and the lens, and the lens and the
source (Schneider, Ehlers, \& Falco 1992).  Loeb \& Perna (1998)
argued that microlensing by stars could therefore be used to resolve the
photospheres of GRB afterglows.  

Recently, Garnavich, Loeb, \& Stanek (2000) have reported the possible
detection of a microlensing magnification feature in the optical-infrared
light curve of GRB 000301C. The achromatic transient feature is well fitted
by a microlensing event of a $0.5 M_\odot$ lens separated by an Einstein
angle from the source center, and resembles the prediction of Loeb \& Perna
(1998) for a ring-like source image with a narrow fractional width ($\sim
10\%$). Alternative interpretations relate the transient achromatic
brightening to a higher density clump into which the fireball propagates
(Berger et al. 2000), or to a refreshment of the decelerating shock either
by a shell which catches up with it from behind or by continuous energy
injection from the source (e.g., Zhang \& M\'esz\'aros 2000).  In order to
regard these alternatives as less plausible, it is essential to identify
the unique temporal and spectral characteristics of a microlensing event.

Gravitational lensing of a point source is achromatic due to the
equivalence principle.  However, differential magnification by an
achromatic lens could still produce a {\it chromatic} magnification signal
if the source is extended and its image looks different at different photon
frequencies (see, e.g. Loeb \& Sasselov 1995). {\it Should a microlensing
event of GRB afterglows be achromatic?} Since the image profile of an
afterglow does depend on frequency (\cite{Sari98}; \cite{PM98}; Granot,
Piran, \& Sari 1999a,b), its magnification lightcurves must exhibit a
predictable level of chromaticity.  In \S 2 we will demonstrate that the
surface brightness profile of an afterglow image changes considerably with
frequency across all spectral breaks in its broken power-law spectrum. We
will therefore show in \S 3 that the magnification history of an
intervening microlens would depend on frequency across spectral breaks.
However, within each power-law segment of the afterglow spectrum, the
magnification would remain achromatic, in rough consistency with the
sparse optical-infrared observations of GRB 000301C.
This spectral behavior offers a fingerprint that can be used to identify a
microlensing event and distinguish it from alternative interpretations.  It
can also be used to constrain the relativistic dynamics of the fireball and
the properties of its gaseous environment, as well as the length scale
required for particle acceleration and magnetic field amplification in
relativistic shocks.

\section{Frequency Dependence of Afterglow Images}

The surface brightness profile (SBP) of an afterglow depends on the
underlying hydrodynamics of the fireball, as well as on the energy
distribution of shock-accelerated electrons. In this {\it Letter} we adopt
the model used by Granot \& Sari (2001, hereafter GS01) to describe
afterglows. Here we provide a brief outline of this model, and refer the
reader to Granot et al. (1999a,b) and GS01 for more details.

The hydrodynamics is described by the Blandford-McKee (1976) self-similar
spherical solution, with a power-law external density profile, $\rho\propto
R^{-k}$, for either $k=0$ or $k=2$, corresponding to a uniform interstellar
medium (ISM) or a stellar wind environment, respectively. The number per
unit energy of accelerated electrons is assumed to be a power law,
$dN_e/d\gamma \propto\gamma^{-p}$ (for $\gamma>\gamma_m$), just behind the
shock.  Thereafter, the initial distribution evolves due to radiative and
adiabatic losses.  The emission mechanism is assumed to be synchrotron
radiation, and the emissivity is integrated over the entire volume behind
the shock front.

The afterglow image is limited to a circle on the sky, whose size grows as
$t^{(5-k)/2(4-k)}$, where $t$ is the observed time. The assumption of a
spherical flow may also serve as an adequate description of a jetted flow,
at sufficiently early times before the jet break time, $t_{jet}$, when the
Lorentz factor of the flow drops below the inverse of the jet opening angle
(Rhoads 1997). In the case of a jet, the image at $t \gtrsim t_{jet}$ is
expected to be different than in the spherical case, and will no longer be
circular for observers who are situated away from the jet axis.  
Deviations from sphericity might also result from hydrodynamic or plasma
instabilities. For simplicity we focus here on a spherical fireball; the
more complex cases mentioned above are left for a future study.

The spectrum of GRB afterglows consists of several power-law segments
(PLSs) where the flux density $F_{\nu}\propto\nu^{\beta}$, which join at
certain break frequencies (\cite{SPN}, hereafter SPN98; \cite{GPS00},
hereafter GPS00; GS01). The cooling time of a typical electron, $t_{cool}$,
is initially smaller than the dynamical time, $t_{dyn}$, resulting in
``fast cooling'' (SPN98), while at latter times $t_{cool}$ becomes larger
than $t_{dyn}$ and there is a transition to ``slow cooling''.  The fast
cooling stage is typically expected to last for the first hour (day) for an
ISM (stellar wind) environment (SPN98; \cite{C&L}).  Since the angular size
of the source is very small at such early times, the source will remain
$\ll\theta_{\rm E}$ and hence unresolved during the fast cooling regime of
a typical microlensing event (or else be resolved only in a negligible
fraction of all events for which the lens nearly coincides with the source
center).  We will therefore restrict our attention to the slow cooling
regime.  The slow cooling spectrum may assume one of three different
shapes, depending on the ordering of the self-absorption frequency,
$\nu_a$, with respect to $\nu_m<\nu_c$, where $\nu_m$ is the typical
synchrotron frequency and $\nu_c$ is the cooling frequency (GS01). The most
common ordering of the break frequencies is the one shown in Figure
\ref{Fig1} with $\nu_a<\nu_m<\nu_c$.  For clarity, we show a broken power
law spectrum, even though in the actual spectrum obtained from our model
the PLSs join smoothly, and the break frequencies are defined where the
asymptotic power laws meet (e.g. Granot et al. 1999b).  Altogether, there
are five different PLSs that appear in the different slow cooling spectra,
which correspond to $\beta=5/2,2,1/3,(1-p)/2,-p/2$.  Due to the
self-similar hydrodynamics, the surface brightness, normalized by its
average value over the image, is independent of time within a given
PLS\footnote{This statement remains valid even when taking into account the
fast cooling spectra, which include two additional PLSs corresponding to
$\beta=11/8,-1/2$ (e.g. SPN98; GPS00), with only one exception:
$\beta=1/3$, where the SBP is different between slow cooling and fast
cooling, since the underlying physics is different.}.  For $\beta=(1-p)/2$
and $\beta=-p/2$, the value of $\beta$ depends on $p$.  The values of
$\beta$ in these two PLSs may therefore accidentally coincide with those of
other PLSs. However, the SBP will still be different, since the underlying
physics is different. Although for these two PLSs the SBP depends on the
value of $p$, we find that at any given time, the differences in the
microlens magnification induced by the differences in the SBP, never exceed
a few percent for $2<p<3$ (and $b\approx 1$, see \S \ref{microlensing}).
Hence, we use a single value of $p=2.5$ in the following.

Figure \ref{Fig2} shows the surface brightness, normalized by its average
value, as a function of the normalized radius inside the image, $r$, for
$k=0,2$ and $\beta=5/2,2,1/3,(1-p)/2,-p/2$. The SBPs for $k=0$ and
$\beta=2,1/3,(1-p)/2$ are taken from Granot et al.  (1999a,b), while for
the other values of $k$ and $\beta$, the SBPs are calculated in a similar
manner, using the formalism of GS01.  For both values of $k$, the image
tends to be more uniform as $\beta$ increases and more ring-like (i.e.  dim
near the center and bright near the outer edge) at low values of $\beta$.
For a given PLS (i.e. a given $\beta$ in Fig. \ref{Fig2}), the image tends
to be more uniform for a stellar wind environment ($k=2$) as compared to an
ISM environment ($k=0$). For $\beta=-p/2$ the surface brightness diverges
at the outer edge of the image ($r=1$), since the emission in this case is
both optically thin and originates from a very thin layer behind the shock,
leading to a relativistic version of the limb brightening effect (e.g.
\cite{Sari98}).  \footnote{The divergence scales as $(1-r)^{-1/2}$, since
the intensity from an optically-thin shell is proportional to its geometric
length along the direction to the observer. This limb brightening effect
does not occur for the other optically-thin PLSs we consider, where the
emission originates from the entire three-dimensional volume behind the
shock. Only the fast-cooling electrons produce the radiation from a very
thin layer just behind the shock front. Note that the diverging cusp makes
a vanishing contribution to the average surface brightness of the afterglow
image.}  However, the physical size of this thin layer cannot be smaller
than that of the transition layer just behind the shock front, where the
electrons are accelerated and the magnetic field is amplified. As will be
shown in \S \ref{microlensing}, the divergence of the SBP at $r=1$ leads to
a very sharp spike in the magnification history.  Introducing a finite
width to the transition layer would eliminate the divergence of the SBP.
However, a thin transition layer would still result in a sharp peak of the
SBP near $r=1$, and a sharp peak in the magnification history, whereas a
thick transition layer will result in a lower and smoother peak in both.
Therefore, the sharpness of the magnification spike may set constraints on
the physical processes of particle acceleration and magnetic field
amplification in relativistic blast waves.

\section{Microlensing Histories}
\label{microlensing}

Next we calculate the magnification histories of GRB afterglows due to
microlensing by an intervening star at a cosmological distance.  Since the
effects of microlensing depend on the angular structure of the image in
units of the Einstein angle $\theta_{\rm E}$ [Eq. (\ref{eq:1})], we
normalize all angular scales by $\theta_{\rm E}$.  The angular radius of
the afterglow image, in units of $\theta_{\rm E}$, is given by
\begin{equation}\label{R_s}
R_s(t)=R_0 t_{days}^{(5-k)/2(4-k)}\ ,
\end{equation}
where $t_{days}$ is the observed time in days and $R_0$ is the angular size
of the image in units of $\theta_{\rm E}$ after one day. The angular
separation between the center of the image and the lens, in units of
$\theta_{\rm E}$, is denoted by $b$. The magnification of a uniform ring of
fractional width $W$ (occupying the interval $1-W<r<1$) is given by
(\cite{LP98}),
\begin{equation}\label{ring}
\mu(R_s,\ W,\ b)={\Psi(R_s,\ b)-(1-W)^2\Psi[(1-W)R_s,\ b]\over
  1-(1-W)^2}
\end{equation}
where $\Psi(R_s,\ b)$ is the magnification of a uniform disk of radius
$R_s$ (Schneider et al. 1992),
\begin{equation}
\Psi(R_s,\ b)={2 \over \pi R_s^2} \left[ \int_{|b-R_s|}^{b+R_s} dR
{{R^2+2}\over{\sqrt{R^2+4}}} {\rm arccos}{{b^2+R^2-R_s^2}\over{2Rb}}+
H(R_s-b){\pi\over2}(R_s-b)\sqrt{(R_s-b)^2+4}\right],
\label{eqn:aus}
\end{equation}
and $H(x)$ is the Heaviside step function.  
In order to obtain the magnification of a radially symmetric image, one can
either divide it into a large number of thin rings, or equivalently use
the resulting integral formula for the frequency-dependent magnification
(\cite{WM94}),
\begin{equation}\label{integral_formula}
\mu_{\nu}(R_s,\ b)=
\int_0^1 2rdr 
\left[\Psi(rR_s,\ b)+{r\over 2} {\partial\Psi\over\partial r}(rR_s,\ 
b)\right]\tilde{I}_{\nu}(r) 
\end{equation}
where $\tilde{I}_{\nu}\equiv I_{\nu}/\langle I_\nu\rangle$ is the specific
intensity normalized by its average value, $\langle I_\nu\rangle=
\int_0^1 2 rdr I_{\nu}(r)$, as shown in Figure \ref{Fig2}.
Note that for a uniform disk ($\tilde{I}_{\nu}(r)\equiv 1$), equation
(\ref{integral_formula}) gives $\mu_\nu=\Psi(R_s,\ b)$, as it should.

The magnification history for $b=1$, $k=0,2$ and
$\beta=5/2,2,1/3,(1-p)/2,-p/2$, is shown in Figure \ref{Fig3}.  The peak in
the magnification history is generally higher and sharper, and occurs at a
slightly earlier time for a smaller spectral slope, $\beta$. This follows
from the general trend of the SBP, which tends to be more sharply peaked
near the outer edge of the image for lower values of $\beta$. At high
values of $\beta$, especially in the optically-thick regime
($\beta=5/2,2$), the magnification peak is relatively shallow and broad.
For an ISM environment ($k=0$), the peak magnification is somewhat higher
and the differences in the magnification histories for different $\beta$
values are more pronounced than for a stellar wind environment
($k=2$). These differences may allow to uncover the density profile of the
circumburst medium from intense monitoring of the microlensing `bump' in an
afterglow light curve. Unfortunately, the quality of the observational data
for GRB 000301C is not sufficient for this purpose.

We now consider the effects of microlensing on the global spectrum.
For clarity, this is illustrated in Figure \ref{Fig1}, on the spectrum
therein. The spectrum consists of several PLSs: $\beta_1=2$,
$\beta_2=1/3$, $\beta_3=(1-p)/2$ and $\beta_4=-p/2$, with different
magnification histories, $\mu_{i}(t)$ ($i=1,...,4$). This causes each
PLS to shift by a different time-dependent factor, as illustrated by
the thin solid line in Figure \ref{Fig1} for $t_{days}=1$. The
different magnification in each PLS generates a shift in the location
of the break frequencies compared to their unlensed values, as
illustrated in the insert at the top right corner of Figure
\ref{Fig1}. The value of the $i$-th break frequency (numbered from low
to high frequencies, as are the PLSs) will shift by a factor of
$\left[\mu_{i+1}(t)/\mu_{i}(t)\right]^{1/(\beta_{i}-\beta_{i+1})}$.
The largest shift occurs at $\nu_c$ since the change in the spectral
slope across the break is rather small ($\beta_{3}-\beta_{4}=1/2$) and
the ratio $\mu_{4}(t)/\mu_{3}(t)$ reaches relatively high values (see
the upper panel of Figure \ref{Fig3}).

\section{Discussion}

We have shown that the frequency dependence of the afterglow image profile
(Fig. 2) introduces a specific level of chromaticity to the magnification
history of GRB afterglows (Fig. 3).  This unique fingerprint can be used to
distinguish between microlensing and alternative hydrodynamic
interpretations for the transient brightening in an afterglow light curve.

\begin{figure}
\begin{center}
\includegraphics[width=15cm]{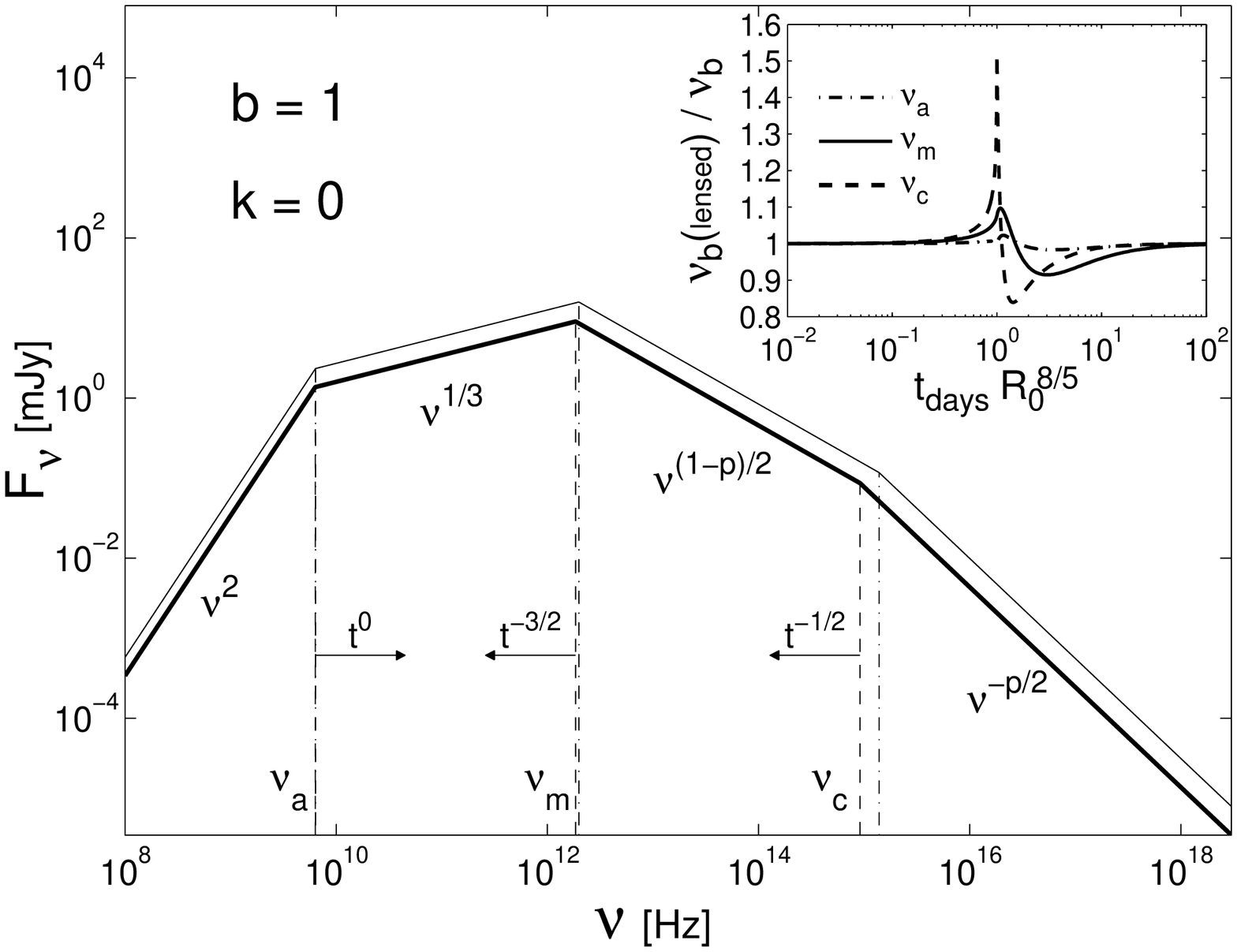} 
\figcaption[fig1]{Typical
    broken power-law spectrum of a GRB afterglow at a redshift $z=1$
    (boldface solid line). The observed flux density, $F_\nu$, as a
    function of frequency, $\nu$, is shown at an observed time
    $t_{days}=1$ for an explosion with a total energy output of
    $10^{52}~{\rm erg}$ in a uniform ISM ($k=0$) with a hydrogen
    density of $1~{\rm cm^{-3}}$, and post-shock energy fractions in
    accelerated electrons and magnetic field of $\epsilon_e=0.1$ and
    $\epsilon_B=0.03$, respectively (using the scalings and notations
    of GS01).  The thin solid line shows the same spectrum, when it is
    microlensed by an intervening star with $b=1$ and $R_0=1$ (see \S
    \ref{microlensing}). The insert shows the excess evolution of the
    break frequencies $\nu_{\rm b}=\nu_a,~\nu_m$ and $\nu_c$ (normalized by
    their unlensed values) due to microlensing.
\label{Fig1}}
\end{center}
\end{figure}

\begin{figure}
\begin{center}
\includegraphics[width=12cm]{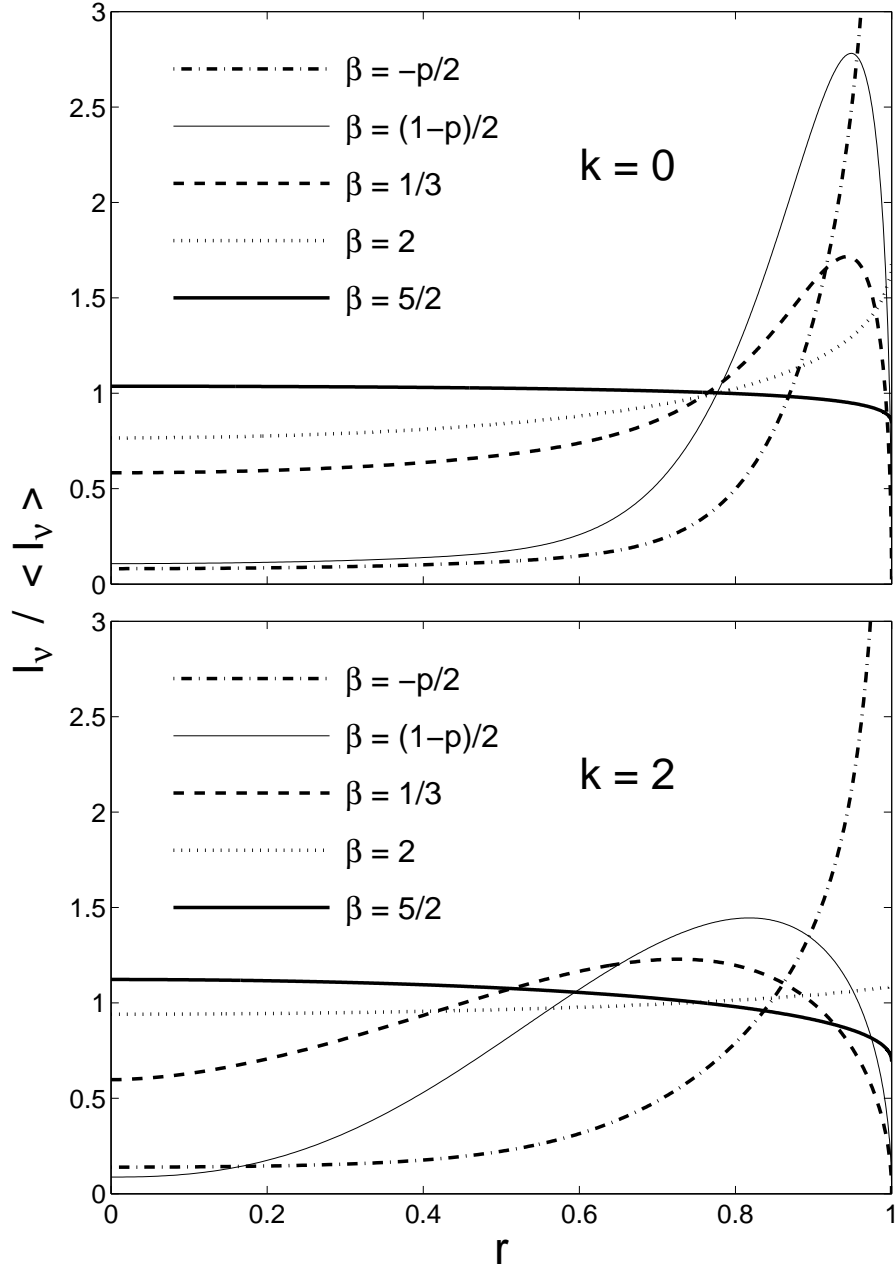}
\figcaption[fig2]{The surface
brightness, normalized by its average value, as a function of the
normalized radius, $r$, from the center of the image (where $r=0$ at the
center and $r=1$ at the outer edge). The image profile changes considerably
between different power-law segments of the afterglow spectrum,
$F_{\nu}\propto\nu^{\beta}$. There is also a strong dependence on the
density profile of the external medium, $\rho\propto R^{-k}$.
\label{Fig2}}
\end{center}
\end{figure}

\begin{figure}
\begin{center}
\includegraphics[width=12cm]{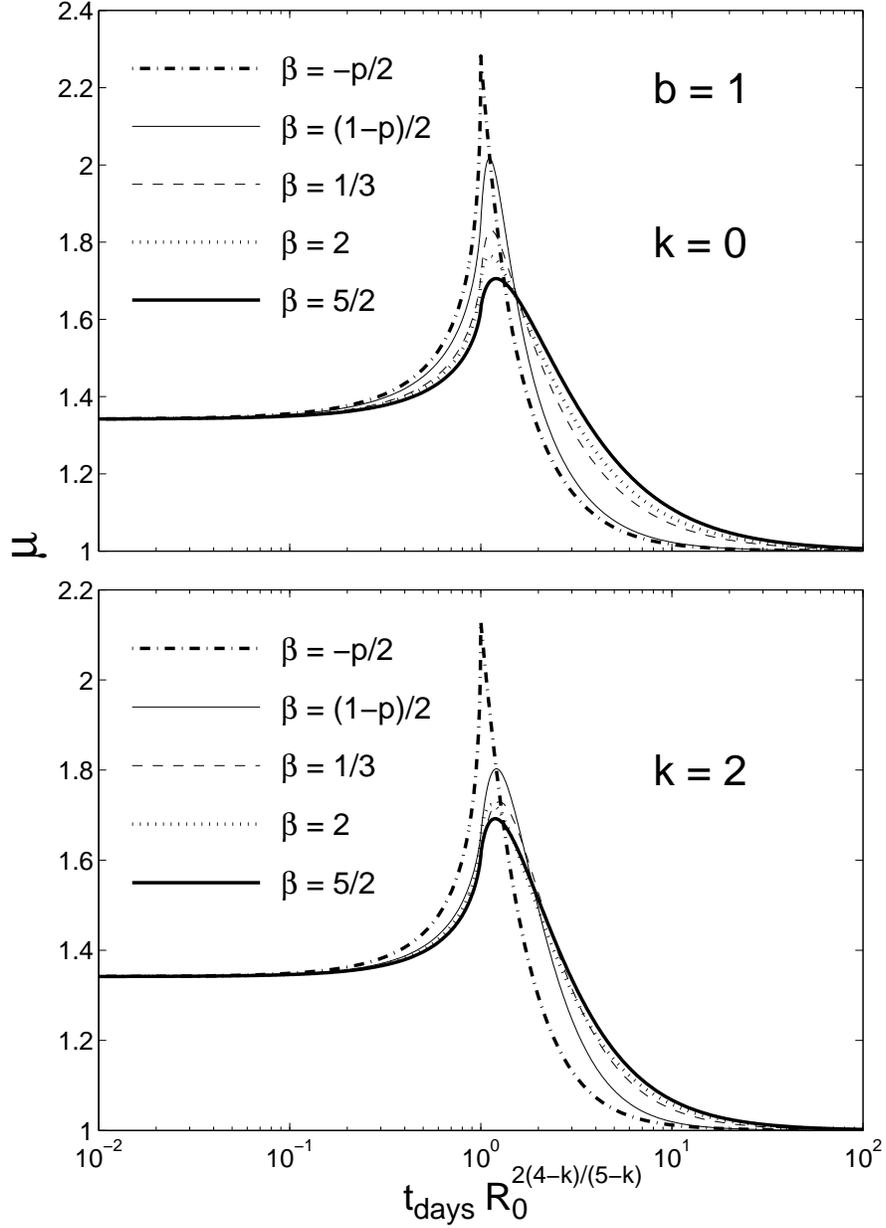}
\figcaption[fig3]{The magnification of the afterglow,
$\mu$, as a function of normalized time, for different power-law segments,
$F_{\nu}\propto\nu^{\beta}$, and different external density profiles
($\rho\propto R^{-k}$ for $k=0,2$). The lens is assumed to be separated by
one Einstein angle from the center of the source ($b=1$).
\label{Fig3}}
\end{center}
\end{figure}

For example, if the GRB blast wave runs into a clump of gas along the
line-of-sight (Berger et al. 2000), then brightening will occur first in
the radio and later in the optical, in opposite order to the microlensing
case. The chromaticity in the clump case is sensitive to the exact geometry
of the clump and the sharpness of its boundary (which could send a reverse
shock into the fireball). In another example, the brightening event may
result from refreshment of the decelerating shock (Zhang \& M\'esz\'aros
2000). In this case, the break frequencies and peak flux of the spectrum in
Figure \ref{Fig1}, namely $\nu_a$, $\nu_m$, $\nu_c$ and $F_{\nu_m}$, will
change by a factor of $f^{1/5}$, $f^{1/2}$, $f^{-1/2}$ and $f$ (or by
$f^{-2/5}$, $f^{1/2}$, $f^{1/2}$ and $f^{1/2}$), respectively, for an ISM
with $k=0$ (or stellar wind with $k=2$) environment, where $f$ is the ratio
between the total energies in the forward shock after and before the
refreshment. This behavior clearly differs from the microlensing prediction
(see insert in Fig. \ref{Fig1}).  By itself, the magnification history of
microlensing is already unique in that it is initially characterized by a
constant positive offset as long as $R_s(t)\ll 1$ (Loeb \& Perna 1998).
Such an initial positive offset does not appear naturally in either of
these alternative interpretations.

Microlensing events are rare but precious. Only one out of roughly a
hundred afterglows is expected to be strongly microlensed (Press \&
Gunn 1973; Blaes \& Webster 1992; Koopmans \& Wambsganss 2000),
although all afterglows are expected to be magnified at some weak
level (Mao \& Loeb 2000).  Detailed monitoring of a few strong
microlensing events among the hundreds of afterglows detected per year
by the forthcoming Swift satellite\footnote{Planned for launch in
  2003; see http://swift.sonoma.edu/}, could be used to constrain the
environment and the dynamics of relativistic GRB fireballs, as well as
their magnetic structure and particle acceleration process (Fig. 3).
Unfortunately, the data on the existing microlensing candidate GRB
000301C is not of sufficient quality to provide firm constraints on
these parameters, although the $R$-band data seems to suggest
$\beta=-p/2$ and gives a somewhat better fit for $k=0$ (a uniform
density environment) than for other parameter choices.  This can be
seen by crudely approximating the SBP as a ring. We find that for
$k=0$, a fraction of $2/3$ ($3/4$) of the emission comes from a ring
of fractional width $W=0.156$ (0.190) for $\beta=(1-p)/2$ and
$W=0.075$ (0.104) for $\beta=-p/2$; while for $k=2$, $W=0.318$ (0.370)
for $\beta=(1-p)/2$ and $W=0.128$ (0.176) for $\beta=-p/2$.  The
$k=0,\beta=-p/2$ values get closest to the empirical best-fit values
of $W=0.07$ using the R-band data (Garnavich et al. 2000).  The
statistical accuracy by which the unknown brightness profile of a
future microlensed afterglow can be reconstructed through an intensive
monitoring campaign, was analyzed in detail recently by Gaudi \& Loeb
(2001).

\acknowledgements

This work was supported in part by the Horowitz foundation and the
US-Israel BSF grant BSF-9800225 (for JG), and by NASA grants NAG
5-7039, NAG 5-7768, NSF grants AST-9900877, AST-0071019 and US-Israel
BSF grant BSF-9800343 (for AL).

\newpage

\newpage

\newpage

\end{document}